\documentstyle[sprocl,amssymb]{article}

\def\Journal#1#2#3#4{{#1} {\bf #2}, #3 (#4)}

\def\PLB{{\em Phys. Lett.}  B}

\def\PRD{{\em Phys. Rev.} D}

\def\be{\begin{equation}}
\def\ee{\end{equation}}
\def\bea{\begin{eqnarray}}
\def\eea{\end{eqnarray}}
\begin{document}
\pagestyle{empty}
\begin{flushright}
{CERN-TH/2000-191}
\end{flushright}
\begin{center}
{\Large \bf SINGLE TOP QUARK \\
           IN THE SM AND BEYOND
\footnote{Presented at 8th International Workshop on
                       Deep-Inelastic Scattering  (DIS'2000), Liverpool, England, 25--30 April 2000}}\\
\vspace*{1cm} 
{\bf     A.        Belyaev    $^{1}$$^{,2}$  
\\
\vspace*{0.3cm} 
          $^1$ CERN Theory Division, \\
	  CH--1211 Geneva 23, Switzerland\\
\vspace*{0.3cm} 	  
          $^2$ Skobeltsyn Institute for Nuclear Physics,\\
               Moscow State University, 119 899, Moscow, Russia}\\
\vspace*{1cm}  
{\bf ABSTRACT} \\ \end{center}
\vspace*{5mm}
\noindent

\vspace*{5cm} 
\noindent 
\rule[.1in]{16.5cm}{.002in}

The prospects of single top quark physics at colliders of the TeV energy
scale are discussed. Single top quark study allows a direct measurement
of the  $Wtb$ vertex and a precise probing physics beyond the Standard
Model in various scenarios.

\noindent
\vspace*{0.5cm}
\begin{flushleft} 
CERN-TH/2000-191\\
June 2000
\end{flushleft}
\vfill\eject
%\pagestyle{empty}
%\clearpage\mbox{}\clearpage

\newpage
\setcounter{page}{1}
\pagestyle{plain}

\section{Introduction}
Single top quark physics has been the subject of  intensive studies for
more than 10 years.  After   the discovery of the top-quark  
it is natural to  study its properties. 
The cross section of the electroweak single top production
(see \cite{sngl-our} and references therein)
is of the same order as that for the strong $t\bar{t}$ pair
production.  Therefore one will have quite enough statistics
for a precise measurement of the electroweak properties of the top quark
involving directly the $Wtb$ vertex.

The top quark is the only known fermion whose
mass is close to the scale of the electroweak symmetry breaking
(EWSB). Hence, the study
of the electroweak properties  of  the top quark sector may shed
light on the mechanism responsible for the EWSB.
Moreover,  deviations from the  SM predictions
might be expected in the large mass top sector.
Thus, single top quark physics could  probe various 
fields of physics beyond the Standard Model:
anomalous gluon--top quark couplings 
(references [65-71] in~\cite{sngl-our}), 
anomalous $Wtb$ couplings 
(\cite{sngl-wtb} and references therein), 
new strong dynamics (\cite{sngl-strong}~and references therein),
flavour changing neutral currents (FCNC) couplings   
(~\cite{sngl-fcnc} and references therein),
 R-parity violating SUSY effects (~\cite{sngl-rpv} and references therein),
CP-violation effects~\cite{sngl-cp},
effects of Kaluza--Klein excited $W$-boson \cite{sngl-extra}.
\section{Hadron colliders}
The main processes of single top quark production at hadron colliders
are:\\
1.~${\mbox{${p\bar{p}}$}}{\mbox{ $\rightarrow$ }}tq{\mbox{${\bar{b}}$}}+X$, 
  ~${\mbox{${p\bar{p}}$}}{\mbox{ $\rightarrow$ }}tq+X$, $W-gluon$ fusion
  process combined with the process involving $b$-quark in the initial state;\\
2.~${\mbox{${p\bar{p}}$}}{\mbox{ $\rightarrow$ }}t{\mbox{${\bar{b}}$}}+X$, 
-- $s$-channel $W^*$ process involving an  off-shell $W$-boson; \\
3.~${\mbox{${p\bar{p}}$}}{\mbox{ $\rightarrow$ }}tW +X$,
where $q$ is a light quark and $X$ represents  the remnants of 
the proton.
These processes are  ordered according to their rates. 
However, it is worth noting  that the relative contributions of $W^*$
and $tW$ processes change with the collider energy. For example,
at the 2 TeV Tevatron collider
$W^*$ process contributes around  30\% to the single top production rate,
while the contribution from the  $tW$ process is only 7\%. The situation at the  14 TeV
LHC collider is just the  opposite.
 
Next-to-leading order (NLO) production rates of the single top at the Tevatron and LHC
are
($m_t=175\mbox{ GeV }, Q^2=m_{top}^2$)~\cite{sngl-nlo}:\\
$\sigma(t\bar{b})=0.88\pm 0.05\mbox{ pb}$,
$\sigma(tq\bar{b}+tq)=2.44\pm 0.4\mbox{ pb}$ for the 2 TeV Tevatron \ \ \ and 
\\
$\sigma(t\bar{b})=10.2\pm 0.6\mbox{ pb}$,
$\sigma(tq\bar{b}+tq)=245\pm 9\mbox{ pb}$ for the 14 TeV LHC collider.
In comparison with complete tree-level calculations~\cite{sngl-our,tait1}, the  NLO results
are some 15\% higher.
The NLO cross section for the $tW$ process is not known yet. The tree-level cross section
of $tW$ production at 2 TeV Tevatron is too small (about 0.2~ pb~\cite{sngl-our}) 
to be seriously considered. However, at the LHC, the  contribution from the $tW$ process to
the total rate of  single top quark production is significant: 
$\sigma(pp\to tW+ \bar{t}W + X)=63^{+16.6}_{-3.6}\mbox{ pb}$~\cite{sngl-tw} 
(CTEQ4L, $Q^2=m_{top}^2$).

It should be stressed that the task of background reduction is a much more serious and
important problem in the case of the single top production than for 
$t\bar{t}$-pair production. This is  because the jet multiplicity of single
top quark events is typically less than  for $t\bar{t}$-pair production and so
QCD $Wjj$ and multijet  backgrounds are much higher.
Therefore the problem of the single top signal extraction is more involved.
The main backgrounds to the single top quark production are: 
$p\bar{p}{\mbox{ $\rightarrow$ }}W+2(3)jets$ 
(gluonic, $\alpha\alpha_s$  order and electroweak $\alpha^2$ order),
$p\bar{p}{\mbox{ $\rightarrow$ }}t\bar{t}$ pair top quark production  and,  
the $j(j)b\bar{b}$ QCD fake background where  one jet
imitates the electron. Initially, the  total background is about of two orders of
magnitude higher than the signal rate but it is possible to work out specific
sets of kinematical cuts~\cite{sngl-back} and  finally get signal-to-background
ratio of $0.4$  and $1.0$ at the Tevatron and LHC respectively. Expected statistics
is about  150 signal events at the Tevatron (for 2 fb$^{-1}$ luminosity)
and $5\times 10^5$ signal events at the LHC (for 100 fb$^{-1}$ luminosity).
With the given statistics one can measure the  $Wtb$ vertex 
with the accuracy of 10\% Tevatron RUN2. In the case of the LHC, the accuracy of the $Wtb$
vertex measurement is basically limited by the uncertainties of parton
distributions, top quark mass and theoretical calculations and expected to be of
the order of few per cent.
\section{Lepton colliders}
The cross section of the SM single top production in 
the $e^+e^-\to e^+\nu_e\bar{t}b$  reaction at LEP2 energies
is of the order of $10^{-5}$--$10^{-6}$ pb,
 which is too small to be observed. However, at TeV  energies,
for example in the $\gamma e \to\nu \bar{t} b$ reaction,
single top quark production rates are  20--80 fb 
for 0.5--2.0 TeV collider energy \cite{sngl-anomlep}.

The measurement of the single top production cross section at
hadron and lepton colliders allows us to put  bounds on the anomalous
$Wtb$ couplings. 
The Lagrangian in unitary gauge, in terms of $F_2^L$ and $F_2^R$ magnetic type 
anomalous couplings,  can be written as:\\
$L = \frac{g}{\sqrt{2}}
    \left[ W^-_\mu \bar{b}\gamma^\mu P_- t
     - \frac{1}{2M_W} W^-_{\mu\nu}
           \bar{b}\sigma^{\mu\nu} (F_2^L P_- + F_2^R P_+) t \right]
     + \mbox{h.\,c.} $,
with $W^\pm_{\mu\nu} = \partial_\mu W^\pm_\nu - \partial_\nu W^\pm_{\mu}$,
$\sigma^{\mu\nu}=i/2[\gamma_\mu,\gamma_\nu]$,
and~$P_\pm=(1\pm\gamma_5)/2$.
The $V+A$ coupling has been skipped since it 
is severely constrained by
the CLEO $b\to s\gamma$ data.

In Table~\ref{sngl-limits} potentials of the Tevatron,
LHC~\cite{sngl-wtb} and  $\gamma e$ \cite{sngl-anomlep} colliders
are compared. One can see that limits on the 
anomalous couplings
that can be obtained at the LHC and 0.5 TeV  $\gamma e$ colliders are comparable,
while the limits for a  2.0 TeV  $\gamma e$ collider are several times better
than those for the LHC.
\begin{table}[h]
  \caption{\label{sngl-limits}
    Uncorrelated limits on anomalous couplings from measurements at
   the Tevatron, LHC and $\gamma e$ colliders}
\begin{tabular}{|l|lcl|lcl|}\hline
        &\multicolumn{3}{|c|}{$F_2^L$}
        &\multicolumn{3}{|c|}{$F_2^R$} \\\hline\hline
      Tevatron ($\Delta_{\mbox{sys.}}\approx10\%$)
               & $-0.18$ &$\ldots$&$+0.55$  & $-0.24$ &$\ldots$&$+0.25$ \\
      LHC ($\Delta_{\mbox{sys.}}\approx5\%$)
               & $-0.052$&$\ldots$&$+0.097$ & $-0.12$ &$\ldots$&$+0.13$ \\
      $\gamma e$ ($\sqrt{s_{e^+e^-}}=0.5\mbox{TeV}$)
               & $-0.1$  &$\ldots$&$+0.1$   & $-0.1$  &$\ldots$&$+0.1$  \\
      $\gamma e$ ($\sqrt{s_{e^+e^-}}=2.0\mbox{TeV}$)
               & $-0.008$&$\ldots$&$+0.035$ & $-0.016$&$\ldots$&$+0.016$ \\
      \hline
 \end{tabular}
\end{table}
\section{FCNC and single top quark production}
Single top quark physics is also a very promising place  to test the 
flavour changing neutral current (FCNC) effects for
$tqV$ couplings,
where $q = u$- or $c$-quark and $V=\gamma,Z,g$.  Those couplings
effectively appear in  supersymmetry or in the scenario where new  dynamics
\cite{fritzsch}
take place in the fermion mass generation. 
The effective Lagrangian involving such  couplings of ($t,q)$ pair 
to massless bosons is the following:\\
$\Delta {\cal L}^{eff} =    \frac{1 }{ \Lambda } \,
[ \kappa_{\gamma} e   \bar t \sigma_{\mu\nu} q
F^{\mu\nu} + \kappa_g g_s \bar t
\sigma_{\mu\nu} \frac{\lambda^i}{2} q G^{i \mu\nu}]
+ \mbox{ h.c.}, $
where $F^{\mu\nu}$ and $G^{i \mu\nu}$ are the usual electromagnetic and gluon
field tensors with  respective FCNC $\kappa_{\gamma}$ and $\kappa_g $
couplings.
It was found \cite{sngl-fcnc} that the 
strength for the anomalous $\bar tcg$
($\bar tug$) coupling may be probed to $\kappa_c / \Lambda =
0.092\rm{~TeV}^{-1}$ ($\kappa_u / \Lambda = 0.026 \rm{~TeV}^{-1}$) at
the Tevatron with $2 \rm{~fb}^{-1}$ of data and $\kappa_c / \Lambda =
0.013\rm{~TeV}^{-1}$ ($\kappa_u / \Lambda = 0.0061 \rm{~TeV}^{-1}$) at
the LHC with $10 \rm{~fb}^{-1}$ of data. Efficiencies from \cite{sngl-fcnc}
can be used to put the limits on $\kappa_{tc\gamma}$, $\kappa_{tu\gamma}$
couplings ($\Lambda=m_{top}$, 95\%CL) at the Tevatron RUN2:
$\kappa_{tc\gamma}<0.24$, $\kappa_{tu\gamma}<0.074$. However, better limits 
(of the order of 0.044 for TeV2) on these couplings $\kappa_\gamma$ are expected to
come from  the study of decays $t \rightarrow q \gamma$ of pair-produced tops. This
process already allows to derive an upper bound $\kappa_\gamma < 0.14$ from the CDF
data taken at  Tevatron RUN1, which is slightly better than the limit obtained by
ALEPH by studying $ee \rightarrow tq$ in LEP2 data.
 
 It is interesting to note that HERA should provide  a very good
sensitivity on FCNC $tu\gamma$ coupling  via single top production.  Even
at the present ZEUS+H1 160 pb$^{-1}$ integrated luminosity, in the absence of
a signal., the limit should be $\kappa_\gamma \lesssim 0.05$, which is
significantly better than the current most stringent bound.
% Assuming a
% total integrated luminosity of 1 fb-1, an upper bound of 0.01 could be
% achieved, which should be competitive with the ultimate TeVatron limit.
    Alternatively, the relatively large ($\sim 1$~pb) 
    cross section still allowed by the current 
    CDF limit on $\kappa_{\gamma,u}$ would lead to many single
    top events. 
    It is interesting to note that H1 observed
    some  events with a high-$P_T$ isolated lepton ($e$ or $\mu$),
    together with missing energy and a large $p_T$ hadronic final
    state. Such a final state would be expected from single
    top events, where the $W$ coming from the top undergoes
    a semileptonic decay\cite{H1MUEV}.
\section{Conclusions}
Single top quark physics is an exciting new window on  possible high mass
scale new physics. 
Single top quark study provides  a direct
measurement of the  $Wtb$ vertex with  10\% accuracy at the Tevatron RUN2 and a few per cent 
accuracy at the LHC and linear colliders. 

In spite of the fact that the present experimental results on the single top search
are consistent with the SM prediction\cite{dudko},
upcoming experiments are very encouraging. With  the new data we can test 
possible deviations from the Standard model such as  anomalous $Wtb$ and FCNC
couplings and probably look into the window beyond the SM.
\section*{Acknowledgments}
The author thanks the Organizing  Committee of DIS'2000 for their warm hospitality and
financial support. He would like to address  special thanks to
John Dainton, Emmanuelle Perez and Magda Lola.
The financial support of CERN Theory Division is gratefully acknowledged.


\section*{References}
\begin{thebibliography}{99}

%\bibitem{sngl-cs} 
%S.D. Willenbrock  and  D. A. Dicus, 		\Journal{\PRD}{34 }{ 155  }{1986}; \ 
%C.-P.~Yuan, 					\Journal{\PRD}{41 }{ 42   }{1990}; \ 
%S. Cortese and R. Petronzio, 			\Journal{\PLB}{253}{ 494  }{1991}; \ 
%G.V. Jikia  and S.R. Slabospitsky, 		\Journal{\PLB}{295}{ 136  }{1992}; \
%R.K. Ellis  and S. Parke, 			\Journal{\PRD}{46 }{3785  }{1992}; \
%D.O. Carlson, E.  Malkawi and C. Yuan,		\Journal{\PLB}{337}{ 145  }{1994}; \ 
%G. Bordes   and B. van Eijk, 			\Journal{\NPB}{435}{  23  }{1995}; \ 
%D.O. Carlson  and  C.P. Yuan,          		MSUHEP-50823 (1995); \ 
%T. Stelzer and S. Willenbrock,          	\Journal{\PLB}{357}{ 125  }{1995}; \
%M.C. Smith  and S. Willenbrock,        		\Journal{\PRD}{ 54}{6696  }{1996}; \
%A.P.~Heinson, A.S. Belyaev and E.E. Boos,	\Journal{\PRD}{ 56}{3114  }{1997}; \
%T. Stelzer, Z. Sullivan and S. Willenbrock,	\Journal{\PRD}{ 56}{5919  }{1997}; \
%T.~Tait and C.-P.~Yuan, 			MSUHEP-71015  (1997)             ; \
%T. Tait, 					\Journal{\PRD}{ 61}{034001}{2000}; \ 
%A. Belyaev  and E. Boos, 			CERN-TH-2000-093(2000).

\bibitem{sngl-our}
A.~Heinson, A. Belyaev and E. Boos,	\Journal{\PRD}{56}{3114}{1997}.

\bibitem{sngl-wtb}
E. Boos, L. Dudko and T. Ohl,   		{\it Eur. Phys. J.} C {\bf 11}, 473(1999).

\bibitem{sngl-strong}
%E.H. Simmons, 					\Journal{\PRD}{ 55}{5494  }{1997};\
%P. Baringer {\it et al.,} 			\Journal{\PRD}{ 56}{2914  }{1997};\
X.-L. Wang  {\it et al.,}			\Journal{\PRD}{60}{014002}{1999}.
%C.-X. Yue and G.-R. Lu,                  	{\it Chin.Phys.Lett.} {\bf 15}, 631 (1998). 

\bibitem{sngl-fcnc}
%E. Malkawi and  T. Tait, 			\Journal{\PRD}{ 54}{5758  }{1996};\
%T. Tait and  C.P. Yuan,  			\Journal{\PRD}{ 55}{7300  }{1997};\ 
%K.J. Abraham, K. Whisnant and B.L. Young,  	\Journal{\PLB}{419}{ 381  }{1998};\
T. Han  {\it et al.,}				\Journal{\PRD}{ 58}{073008}{1998}.
%T. Han and  J. L. Hewett,                   	\Journal{\PRD}{ 60}{074015}{1999};\
%S. Bar-Shalom and  J. Wudka,                	\Journal{\PRD}{ 60}{094016}{1999}.

\bibitem{sngl-rpv}
%A. Datta {\it et al.,}				\Journal{\PRD}{ 56}{3107}{1997};\
%J. Oakes {\it et al.,}				\Journal{\PRD}{ 57}{534 }{1998};\
%E.L. Berger, B.W. Harris and  Z. Sullivan,	\Journal{\PRL}{ 83}{4472}{1999};\
P. Chiappetta {\it et al.,}			\Journal{\PRD}{ 61}{115008}{2000}.
%M. Chemtoband  G. Moreau,			\Journal{\PRD}{ 61}{116004}{2000}.

\bibitem{sngl-cp}
S.~Bar-Shalom, D.~Atwood and A.~Soni, 		\Journal{\PRD}{ 57}{1495}{1998}.

\bibitem{sngl-extra}
A.~Datta {\it et al.,}				\Journal{\PLB}{483}{203}{2000}. 

\bibitem{sngl-nlo}
M.C. Smith  and S. Willenbrock,        	     	\Journal{\PRD}{ 54}{6696 }{1996}; \
T. Stelzer, Z. Sullivan and S. Willenbrock, 	\Journal{\PRD}{ 56}{5919 }{1997}.

\bibitem{tait1} 
T.~Tait and C.-P.~Yuan, 			MSUHEP-71015  (1997), { \sf hep-ph/9710372}            ; \

\bibitem{sngl-tw} 
T. Tait, 					\Journal{\PRD}{ 61}{034001}{2000}; \ 
A. Belyaev  and E. Boos, 			CERN-TH-2000-093(2000), {\sf hep-ph/0003260}.



\bibitem{sngl-back} 
A.~Belyaev, E.~Boos and L.~Dudko,        \Journal{\PRD}{ 59}{075001}{1999}.

%\bibitem{sngl-lep} 
%E. Boos {\it et al.,}				\Journal{\PLB}{326}{190}{1994}.

\bibitem{sngl-anomlep} 
E. Boos {\it et al.,}        			\Journal{\PLB}{404}{119}{1997}. 

%\bibitem{whisnant}
%K.~Whisnant, J.~Yang, B.~Young and X.~Zhang,    \Journal{\PRD}{ 56}{467}{1996}.

\bibitem{fritzsch}
H. Fritzsch,				       \Journal{\PLB}{184}{391}{1987}.


%\bibitem{fcnc-cdf}
%F. Abe et al., CDF Collaboration                \Journal{\PRL}{80}{2525}{1998}.

\bibitem{H1MUEV}
 H1 Collaboration, C.~Adloff {\it et al.},
 {\it Euro. Phys. J. } C{\bf 5} 575 (1998); N.~Malden, these proc.

\bibitem{dudko}
L. Dudko for D\O \  and CDF collaborations, preliminary results presented at the 
 XXXVth Rencontres de Moriond (QCD), March 2000.


\end{thebibliography}
\end{document}